\documentclass[11pt, oneside,onecolumn]{article}   	

\usepackage{geometry}  
\usepackage{algorithm} 
\usepackage[noend]{algpseudocode}
\usepackage{graphicx}				
\usepackage{amssymb}
\usepackage{amsmath}
\usepackage{amsfonts}
\usepackage{hyperref}
\usepackage{color}
\usepackage{xcolor}
\usepackage{authblk}
\usepackage{amsthm}
\usepackage{amssymb}
\usepackage{url}
\usepackage[colorinlistoftodos,prependcaption,textsize=tiny]{todonotes}
\usepackage{regexpatch}
\makeatletter
\xpatchcmd{\@todo}{\setkeys{todonotes}{#1}}{\setkeys{todonotes}{inline,#1}}{}{}
\makeatother

\usepackage[utf8]{inputenc} 
\usepackage{textcomp} 
\usepackage{graphicx}  
\usepackage{flafter}  

\usepackage{amsmath,amssymb}  
\usepackage{bm}  

\usepackage{mdframed}

\theoremstyle{plain} 
\newtheorem{thm}{\protect\theoremname}[section]

 \theoremstyle{plain}

\theoremstyle{plain}

  \providecommand{\definitionname}{Definition}
  \providecommand{\lemmaname}{Lemma}
    
  \providecommand{\theoremname}{Theorem}
\providecommand{\theoremname}{Theorem}

\begin{document}


\title{Beyond pairwise network similarity: exploring Mediation and Suppression between networks}

\author[1,2]{Lucas Lacasa}
\author[3,4]{Sebastiano Stramaglia}
\author[5,6]{Daniele Marinazzo}

\affil[1]{School of Mathematical Sciences, Queen Mary University of London, E14NS London (UK)}
\affil[2]{Institute for Cross-Disciplinary Physics and Complex Systems (CSIC-UIB), Mallorca (Spain)}
\affil[3]{Dipartimento Interateneo di Fisica, Università Degli Studi di Bari, Aldo Moro, 70126 Bari (Italy)}
\affil[4]{INFN, Sezione di Bari, 70126 Bari (Italy)}
\affil[5]{Data Analysis Department, Ghent University, 9000 Ghent (Belgium)}
\affil[6]{IRCCS Ospedale San Camillo s.r.l., 30126
Venice (Italy)}
\date{l.lacasa@qmul.ac.uk, sebastiano.stramaglia@ba.infn.it, daniele.marinazzo@ugent.be}							

\maketitle

\abstract{Network similarity measures quantify how and when two networks are symmetrically related, including measures of statistical association such as pairwise distance or other correlation measures between networks or between the layers of a multiplex network, but neither can directly unveil whether there are hidden confounding network factors nor can they estimate when such correlation is underpinned by a causal relation.
In this work we extend this pairwise conceptual framework to triplets of networks and quantify how and when a network is related to a second network directly or via the indirect mediation or interaction with a third network. Accordingly, we develop a simple and intuitive set-theoretic approach to quantify mediation and suppression between networks. We validate our theory with synthetic models and further apply it to triplets of real-world networks, unveiling mediation and suppression effects which emerge when considering different modes of interaction in online social networks and different routes of information processing in the brain.}



\section{Introduction}
Networks are usually seen as a parsimonious model to describe the backbone architecture of complex systems \cite{latora}. Accordingly, comparing different systems boils down to compare their architecture, leading to the notion of network similarity measure \cite{distance0,distance1,distance2,distance3,distance4}. In graph theory, two graphs are isomorphic if there exist a vertex permutation which maps one network into the other, naturally leading to a binary (and not very useful in real-world systems) notion of similarity. More useful approaches proceed by projecting networks into a suite of properties summarised in some vector ${\bf p}$ (e.g. degree distribution, centrality vectors, eigenspectra, etc) and, subsequently construct a similarity metric $\cal D$ by which two networks $A$ and $B$ are closer in the space spanned by $\bf p$ if ${\cal D}(A,B)=||{\bf p}_A - {\bf p}_B||$ is ``small''. Other ideas include the formalisation of graph kernels \cite{distance0}, comparing networks by comparing the statistics of random walks running over them \cite{distance5, distance6}, or using statistical approaches such as estimating topological correlations between networks. While in all these approaches we typically have ${\cal D}(A,B)={\cal D}(B,A)$, i.e. a symmetrical relation, in many cases this undirected relation is hiding an actual direction (whether causal or not). As an example, consider social networks. The different layers of the social network of an individual are typically correlated: my friends offline tend to be also friends in Facebook. However, such {\it relation} is directional: when a new link --i.e, a new social relationship-- is created, then it is likely that such a link will be replicated within her online social network too (Facebook, Instagram), but it is proportionally less likely that the direction of influence is inverted. So the offline and the online social network of a person are probably similar, but such similarity has a direction. 
Furthermore, in many cases such influence is not direct (not causal). Sometimes, there is a hidden network $\cal C$ that indeed confounds or mediates the relation between $\cal A$ and $\cal B$. For instance, the Facebook and Instagram networks of a certain individual are correlated not because there is a direct, causal relationship between them, but because both these networks are indeed related to the actual (offline) social network of the person.\\

In this work we are interested in understanding and disambiguating when the {relation} between two networks $\cal A$ and $\cal B$ (where for instance ${\cal A}\ r \ {\cal B}$ if ${\cal D}({\cal A},{\cal B})<\epsilon$) is a direct one or is underpinned by the hidden interaction with a third network $\cal C$. 
In particular, $\cal C$ can be independent of $\cal A$ and $\cal B$ (leading to a direct relation ${\cal A}\ r \ {\cal B}$). $\cal C$ can also act as a hidden {\it mediator} or {\it confounding} factor (${\cal A}\ r\ {\cal C}$, ${\cal C}\  r\ {\cal B}\Rightarrow {\cal A}\ r\ {\cal B})$. Finally, ${\cal C}$ can act as a suppressor such that $[{\cal A}\oplus {\cal C}]\ r\ {\cal B}$, where $\oplus$ is here to be defined but conceptually means $\cal{A}$ and $\cal C$ interact synergistically. The terms {\it mediator}, {\it confounder} and {\it suppressor} are inspired by the information-theoretic framework described in \cite{mackinnon}.\\
In what follows we address these questions introducing a set-theoretical approach where concepts such as network mediation or network suppression emerge naturally. We benchmark our theory with simple generative models and then apply it to a range of empirical networks, where we unveil and discuss the concomitant roles of mediation and suppression. 
\\

\section{Theory}

Let $\mathcal{A, B, C}$ be three unweighted networks with adjacency matrices $\bf A, B, C$, all with the same node set and respective edge sets $a,b$ and $c$ (i.e. they can also be identified with the layers of a multiplex network). Let us define the network-Jaccard index of two networks $\text{NJ}({\mathcal A,B})$ as the Jaccard index over its edge sets 
\begin{equation}
\text{NJ}({\cal A,B}):= J(a,b)=
\frac{|a \cap b|}{|a \cup b|}
\label{eq:Jaccard}
\end{equation}
$\text{NJ}({\mathcal A,B})$ is a similarity metric, and a distance can be easily defined as $d({\mathcal A,B})=1-\text{NJ}({\cal A,B})$. This quantity alone can be used to initially establish if two networks are related.
Regardless the fact that such relation is effectively undirected or otherwise is causal (influence), in order to explore whether such relation is underpinned by a third network $\cal C$ we need to quantify the effect of conditioning such relation on $\cal C$. Let us then define the partial network-Jaccard index $\text{NJ}_p(\mathcal{A,B}|\mathcal{C})$ of two networks $\mathcal{A, B}$ conditioned on a third one $\mathcal{C}$ as the Jaccard index over the edge subsets of $\mathcal{A}$ and $\mathcal{B}$ formed by those edges which are absent in $\mathcal{C}$:
\begin{equation}
\text{NJ}_p(\mathcal{A,B}|\mathcal{C}) = \frac{|(a\cap b) \setminus c|}{|(a\cup b) \setminus c|}
\label{eq:PartJaccard}
\end{equation}
Let's see intuitively the effect of conditioning with respect to $\mathcal{C}$ in this way.
Suppose initially that $\mathcal{C}$ is totally independent from $\mathcal{A}$ and $\mathcal{B}$. Then we may expect that the Jaccard index, on average, will be the same if evaluated just on the links which are absent in $\mathcal{C}$, so
$\text{NJ}_p(\mathcal{A,B}|\mathcal{C})\approx \text{NJ}({\cal A,B})$. Suppose on the other hand that $\mathcal{A}$ is influencing $\mathcal{B}$ indirectly, with the mediation of $\mathcal{C}$. Then, intuitively, removing the links of $\mathcal{C}$ would effectively push the partial Jaccard index to zero.
A similar scenario takes place if $\mathcal{A}$ and $\mathcal{B}$ are undirectedly related through direct relation to a confounding factor $\mathcal{C}$.
Finally, $\mathcal{C}$ could be suppressing the influence of $\mathcal{A}$ in $\mathcal{B}$. For example, imagine that $\mathcal{B}$ somewhat depends on whether $\mathcal{A}$ and $\mathcal{C}$ interact synergistically, e.g. if links in $\mathcal{C}$ are more likely to occur if they are in one network but not on the other (probabilistic XOR gate); then removing the links of $\mathcal{C}$ will enhance the partial Jaccard index.\\
To distinguish these three scenarios, we define the Jaccard net difference
\begin{equation}
\Delta[\mathcal{A,B; C}] :=\Delta= \text{NJ}_p(\mathcal{A,B}|\mathcal{C}) - \text{NJ}({\cal A,B}).
\end{equation}
Intuitively, if ${\cal C}$ is independent of the relation between ${\cal A}$ and ${\cal B}$ then $\Delta \approx 0$, if it mediates or confounds such relation then $\Delta <0$, and if it acts as a suppressor then $\Delta >0$.\\ 
\noindent In what follows we construct simple generative models of independent, mediated and suppressor interactions, detailed as algorithms, we prove that these correctly generate these three types of trivariate relations, and depict numerical simulations of the outcome for finite networks.
\begin{figure}[htb!]
   \centering
   \includegraphics[width=0.75\textwidth]{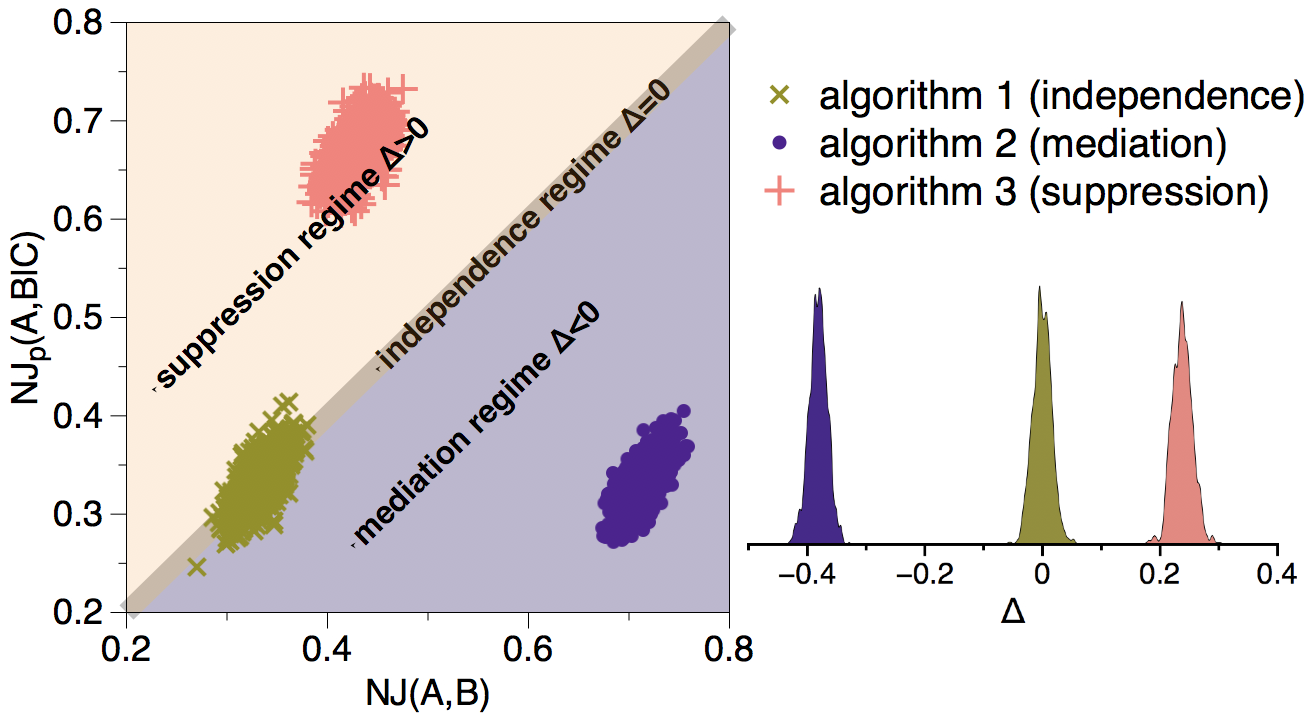} 
   \caption{{\bf Pure models. }$\text{NJ}_p(\mathcal{A,B}|\mathcal{C})$ vs $\text{NJ}({\cal A,B})$, calculated on $1000$ realizations of triplets of networks of $N=50$ nodes wired such that $\mathcal{C}$ plays no effect (green crosses), plays a mediating effect (violet dots) or a suppression effect (red crosses) in the relation between $\mathcal{A}$ and $\mathcal{B}$. These interactions are constructed using the generative models described in Algorithms 1, 2 and 3 ($p=0.5$ in every case, and $q=1$). For completeness, we depict the histograms $P(\Delta)$ which certify that these algorithms generate networks where $\cal C$ play an independent role ($\Delta \approx 0$), a mediating role ($\Delta < 0$) or a suppressing role ($\Delta >0$).}
   \label{fig:toy_numerics}
\end{figure}


\subsection{Independency}

A simple generative model of independency is given by three independent, Erdos-Renyi-type models, where in each of the networks each possible link independently occurs with probability $p$, see Algorithm 1.
\begin{algorithm}[h!]
\caption{{\textsc{Uncorrelated}()}}\label{euclid}
\begin{flushleft}
\textbf{Output}:  3 Erdos-Renyi adjacency matrices $\bf A, B, C$ which are uncorrelated ($\Delta \approx 0)$
\end{flushleft}
\vspace{-4mm}
\begin{algorithmic}[1]
\State ${\bf A} \gets {\bf 0}$ 
\State ${\bf B} \gets {\bf 0}$ 
\State ${\bf C} \gets {\bf 0}$ 

\For{\texttt{$i=1 \ {\bf to} \ N$}}
       \For{\texttt{$j=i+1 \ {\bf to} \ N$}}
         \If{\textsc{rand} $<p$} $A_{ij}, A_{ji} \gets 1$ \EndIf
         \If{\textsc{rand} $<p$} $B_{ij}, B_{ji} \gets 1$ \EndIf
          \If{\textsc{rand} $<p$} $C_{ij}, C_{ji} \gets 1$ \EndIf
\EndFor 
\EndFor 
\State \Return {\bf A,B,C}
\end{algorithmic}
\vspace{3mm}
  \end{algorithm}
The following theorem can be easily proved.  
\begin{thm}
\label{thm:uncorr}
Let $\bf A$, $\bf B$ and $\bf C$ be as in Algorithm 1. Then $\mathbb{E}(\Delta)=0$ and the expected values of $\text{NJ}_p({\cal A,B}|{\cal C})$ and $\text{NJ}({\cal A,B})$ are equal to $p/(2-p)$.
\end{thm}
\noindent The proof of this theorem is given in the appendix. We conclude that, on average, $\text{NJ}_p$ and $\text{NJ}$ are nearly equal for uncorrelated networks generated in this way, as partialization with respect to an independent network $\mathcal C$ does not have any effect. In Fig.\ref{fig:toy_numerics} we illustrate this case for finite networks with $N=50$ nodes and $p=0.5$, finding that indeed $\Delta \approx 0$ and that $\text{NJ}_p({\cal A,B}|{\cal C})\approx \text{NJ}({\cal A,B}) \approx 1/3$, in good agreement with the theorem.

\begin{figure}[!htb]
   \centering
   \includegraphics[width=0.55\textwidth]{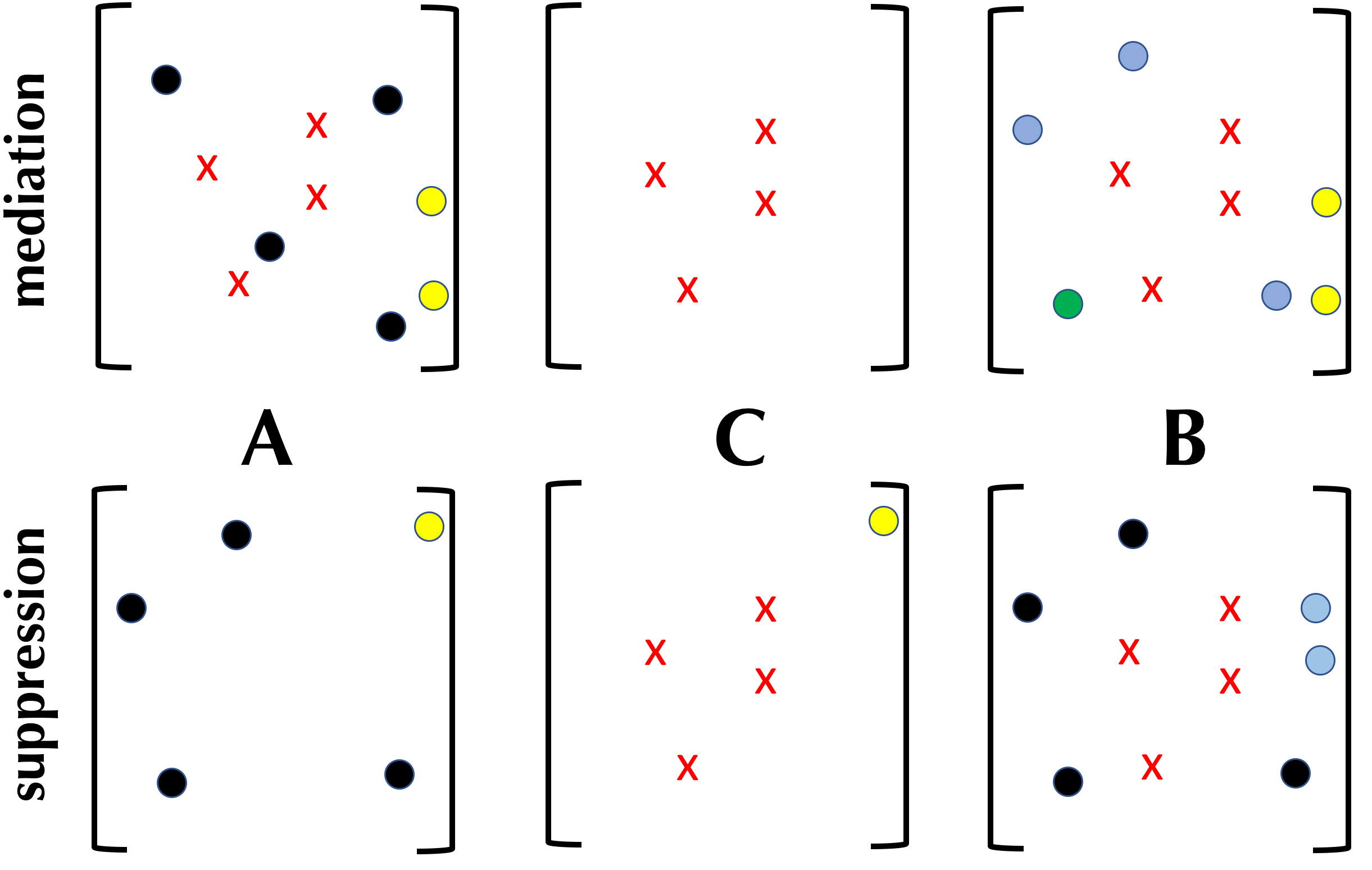} 
   \caption{{\bf Adjacency matrices cartoons displaying mediation and suppression.} (Top) Network $\cal C$ is mediating the relation between $\cal A$ and $\cal B$ (Algorithm 2). (Bottom) Network $\cal C$ acts as a suppressor between $\cal A$ and $\cal B$ (Algorithm 3).}
  \label{fig:matrices}
\end{figure}

\subsection{Mediation}
Suppose now that $\mathcal{A}$ and $\mathcal{B}$ are both dependent on $\mathcal{C}$, i.e. where there is a link in $\mathcal{C}$, then there is a link in $\mathcal{A}$ and $\mathcal{B}$ (see Fig.\ref{fig:matrices} for an illustration of such case, and Algorithm 2 for a formal recipe of this generative model). 

\begin{algorithm}
\caption{{\textsc{Mediated}()}}\label{euclid}
\begin{flushleft}
\textbf{Output}:  3 Erdos-Renyi adjacency matrices $\bf A, B, C$ where $\bf C$ mediates relation between $\bf A$ and $\bf B$ ($\Delta > 0)$
\end{flushleft}
\vspace{-4mm}
\begin{algorithmic}[1]
\State ${\bf A} \gets {\bf 0}$ 
\State ${\bf B} \gets {\bf 0}$ 
\State ${\bf C} \gets {\bf 0}$ 

\For{\texttt{$i=1 \ {\bf to} \ N$}}
       \For{\texttt{$j=i+1 \ {\bf to} \ N$}}
         \If{\textsc{rand} $<p$} $A_{ij}, A_{ji} \gets 1$ \EndIf
         \If{\textsc{rand} $<p$} $B_{ij}, B_{ji} \gets 1$ \EndIf
          \If{\textsc{rand} $<p$} $C_{ij}, C_{ji},A_{ij}, A_{ji},B_{ij}, B_{ji}  \gets 1$ \EndIf
\EndFor 
\EndFor 
\State \Return {\bf A,B,C}
\end{algorithmic}
\vspace{3mm}
  \end{algorithm}

\noindent This describes a situation where $\mathcal{C}$ mediates the relation between $\mathcal{A}$ and $\mathcal{B}$ (or, alternatively, $\mathcal{C}$ is confounding that relation). Partializing with respect to $\cal C$ removes the dependence between ${\cal A}$ and ${\cal B}$ due to $\cal C$, which intuitively leads to $\Delta<0$.
The following theorem can be proved:
\begin{thm}
\label{thm:mediated}
Let $\bf A$, $\bf B$ and $\bf C$ be as in Algorithm 2. If $\bf A$ and $\bf B$ share at least one edge besides the common edges shared with $\bf C$, then $\Delta<0$.
\end{thm}
\noindent The proof of this theorem is also put in an appendix. In Fig.\ref{fig:toy_numerics} we show numerical results for finite networks with $N=50$, with $p=0.5$.\\

\subsection{Suppression}
Finally, let us consider the case where $\cal B$ depends on the interaction of $\cal A$ and $\cal C$ such that, an edge occurs in $\cal B$ with a certain probability if it appears in $\cal A$ but not in $\cal C$ or alternatively if it appears in $\cal C$ but not in $\cal A$ (see Fig. \ref{fig:matrices} for an illustration). This is akin to a probabilistic XOR gate.
Then on average $\text{NJ}_p(\mathcal{A,B}|\mathcal{C})> \text{NJ}(\mathcal{A,B})$, i.e. partializing with respect to $\cal C$ in this case evidences suppression effects.\\

\begin{algorithm}
\caption{{\textsc{Suppression}()}}\label{euclid}
\begin{flushleft}
\textbf{Output}:  3 Erdos-Renyi adjacency matrices $\bf A, B, C$ where $\bf C$ acts as a suppressor between $\bf A$ and  $\bf B$ ($\Delta > 0)$
\end{flushleft}
\vspace{-4mm}
\begin{algorithmic}[1]
\State ${\bf A} \gets {\bf 0}$ 
\State ${\bf B} \gets {\bf 0}$ 
\State ${\bf C} \gets {\bf 0}$ 

\For{\texttt{$i=1 \ {\bf to} \ N$}}
       \For{\texttt{$j=i+1 \ {\bf to} \ N$}}
         \If{\textsc{rand} $<p$} $A_{ij}, A_{ji} \gets 1$ \EndIf
         \If{\textsc{rand} $<p$} $B_{ij}, B_{ji} \gets 1$ \EndIf
          \If{\textsc{rand} $<p$} $C_{ij}, C_{ji} \gets 1$ \EndIf
          \If{$A_{ij}*C_{ij}=0$ \& $A_{ij} + C_{ij}>0$ \& \textsc{rand} $<q$} $B_{ij}, B_{ji} \gets 1$ \EndIf
\EndFor 
\EndFor 
\State \Return {\bf A,B,C}
\end{algorithmic}
\vspace{3mm}
  \end{algorithm}
  
\noindent Algorithm 3 encapsulates the generative model. The following theorem can be proved:
  \begin{thm}
Let $\bf A$, $\bf B$ and $\bf C$ be as in Algorithm 3. Then $\mathbb{E}(\Delta)>0$.
\end{thm}

\noindent The proof for this theorem is also put in an appendix. In Fig.\ref{fig:toy_numerics} we show numerical results for finite networks with $N=50$, with $p=0.5$ and $q=1$, which are in full agreement with the theorem.

\begin{figure*}[!htb]
   \centering
\includegraphics[width=0.48\textwidth]{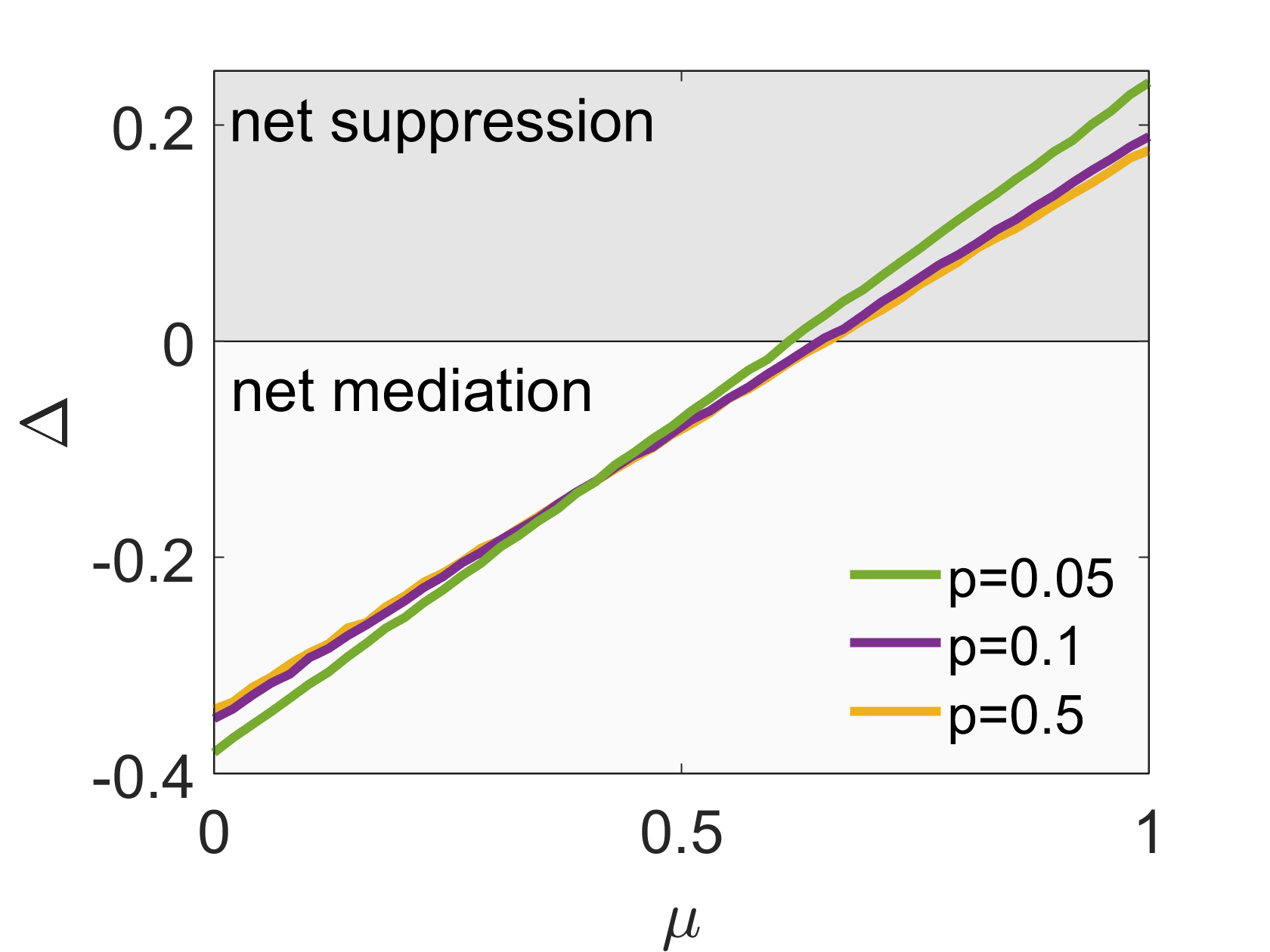} 
\includegraphics[width=0.48\textwidth]{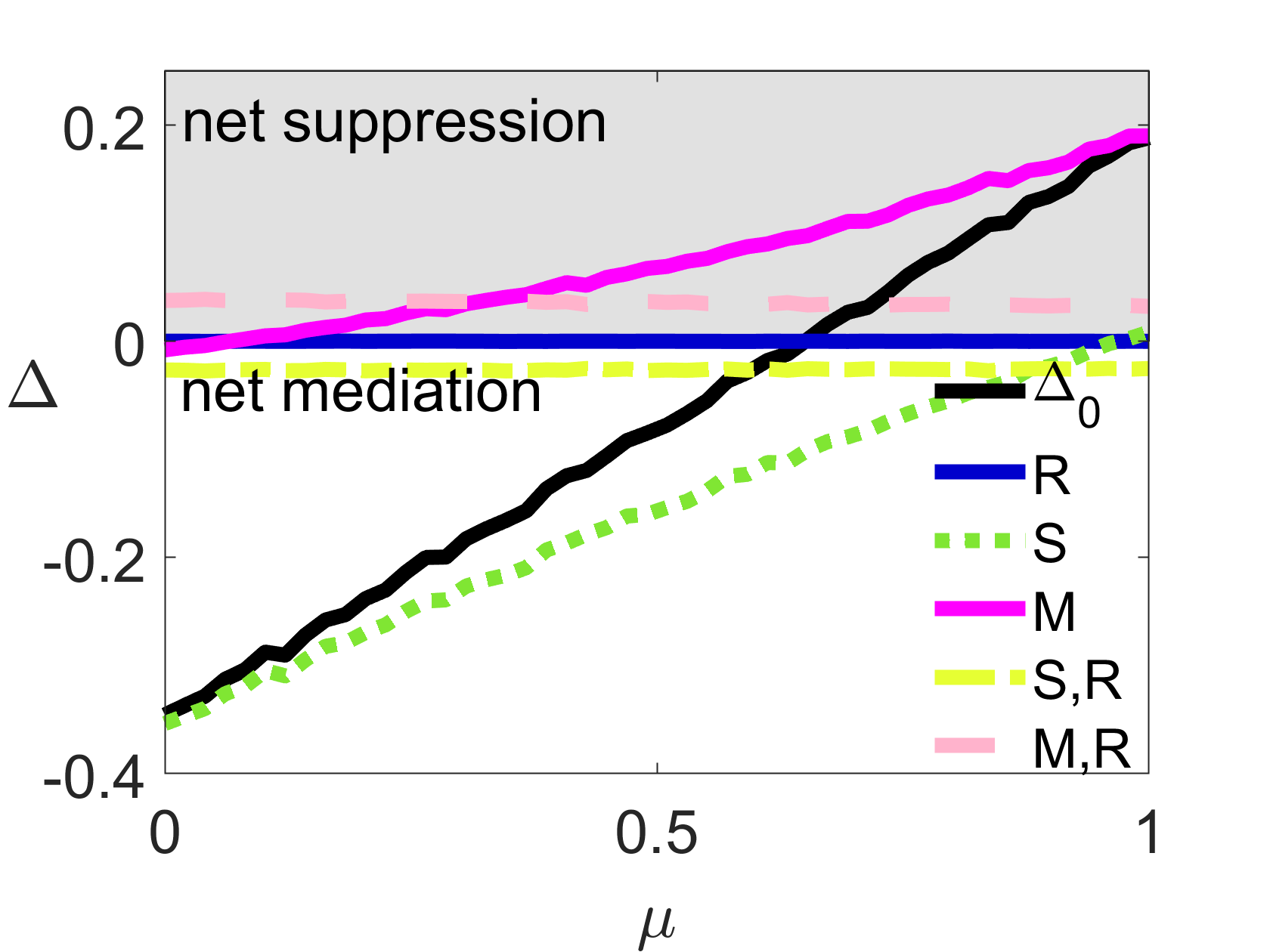} 
   \caption{(Left panel) Values of $\Delta$ obtained averaging 50 realizations of an interpolating model giving a blend of mediation and suppression (see the text), where each network has $N=300$ nodes and varying model parameter $p$, as a function of the interpolation parameter $\mu$ (in each of the N(N-1) steps we apply Algorithm 2 with probability $1-\mu$ and Algorithm 3 with $\mu$). 
  (Right panel) Revised version of the  left panel, this time selectively removing all suppression and mediation effects according to our theoretical framework. The original curve, for an interpolating model generating a blend of mediation and suppression, is depicted in black. The blue curve is a pure randomisation, which generates $\Delta\approx 0$. The dotted green line corresponds to a selective rewiring that removes all hidden  suppression: in  that case the curve is kept always below zero (increasing $\mu$ increases the amount of Algorithm 3, but then is selectively rewired, thus effectively randomising the networks and pushing $\Delta$ to zero). The  pink curve is the result of a selectively rewiring  that removes all hidden mediation: in that case the curve  is pushed to the regime $\Delta >0$. As $\mu$ increases, the amount of Algorithm 3 (generating suppression) is increased, hence pushing $\Delta$ to larger values. The dashed yellow and pink lines are the result of selectively rewiring on the randomised networks, and only highlight the residual values of suppression or mediation which occur by chance (as a finite size effect) in randomised networks.}
   \label{fig:interpol}
\end{figure*}

\subsection{Coexistence of mediation and suppression effects}
When we abandon ideal cases where only suppression or only mediation are present, and we go towards a mixture of the two, it becomes evident that both effects can be hidden and a single $\Delta$ cannot in principle tell us if the system evidences {\it only} one out of two mechanisms. To investigate coexistence of both mechanisms, we run a simulation in which algorithms 2 and 3 above are combined: in each step with probability $1-\mu$ we take algorithm 2 and with probability $\mu$ we take algorithm 3. The resulting model linearly interpolates mediation and suppression, such that measurable $\Delta = (1-\mu) \Delta_{\text{med}} + \mu\Delta_{\text{syn}}$, where $\Delta_{\text{med}}$ and $\Delta_{\text{syn}}$ are hidden.
The results are depicted in the left panel of Fig.\ref{fig:interpol}, for different instances of parameter $p$ and $q=1$. We can have negative, null or positive values of $\Delta$ underpinned by a balance of both mediation and suppression mechanisms, and actually for the concrete set of parameters,
the effect of mediation in $\Delta$ is slightly stronger than the effect of suppression (this unbalance gets more pronounced for $q<1$). 
This simple interpolating model thus leads us to conclude that, in real cases, we might for instance be naively measuring $\Delta <0$ and misleadingly concluding that there is only mediation where in fact both mediation and suppression could be at play. Accordingly, a measure describing the effects of suppression and mediation is not enough to describe and resolve the simultaneous presence of both.\\

\begin{algorithm}[h!]
\caption{{\textsc{Null}()}}\label{Algo4}
\begin{flushleft}
{\bf Input}: 3 adjacency matrices {\bf A, B, C} and mode $X$ (mediation (M) or suppression (S))\\
{\bf Output}:  Null model net difference $\Delta_{W,X}$
\end{flushleft}
\vspace{-3mm}
\begin{algorithmic}[1]
\State ${\bf B}2 \gets \textsc{FullRewire}({\bf A, B, C})$
\State $\Delta_X \gets \textsc{SelectiveRewire}({\bf A,B,C})$
\State $\Delta_{RX} \gets \textsc{SelectiveRewire}({\bf A,B2,C}) $
\State $\Delta_{W,X} \gets [\Delta_X - \Delta_{RX}]/X_{\text{max}}$ 
\State \Return ${\Delta_{W,X}}$
\end{algorithmic}
\vspace{3mm}
\begin{algorithmic}[1]
    \Function{\textsc{FullRewire}}{$\bf G$}
 \For{$\ell_{ij} \in {\bf G}$}
         \If{\textsc{rand} $<p$} $\ell_{ij} \gets \ell_{kl}$,  $\ell_{kl} \gets \ell_{ij}$
         \EndIf
\EndFor 
  \State  \Return {\bf G}
    \EndFunction
  \end{algorithmic}
  \vspace{3mm}
  \begin{algorithmic}[1]
    \Function{\textsc{SelectiveRewire}}{{\bf A, G, C}, X}
 \If{$X=S$}
 \For{$\ell'_{ij} \in {\bf G}'$}
         \If{$\ell'_{ij}=1 \ \& \ A_{ij}=1\  \& \ C_{ij}\neq1$} $\ell'_{kl} \gets \ell'_{ij}; \ \ell'_{ij} \gets 0 $ 
         \EndIf
 \EndFor
 \EndIf
 \If{$X=M$}
  \For{$\ell'_{ij} \in {\bf G}'$}
         \If{$\ell'_{ij}=1 \ \& \ A_{ij}=1\  \& \ C_{ij}=1$} $\ell'_{kl} \gets \ell'_{ij}; \ \ell'_{ij} \gets 0 $ 
         \EndIf
 \EndFor
 \EndIf
\State $\textsc{selective-rewire} \gets \text{NJ}_p({\bf A,G}'|{\bf C}) - \text{NJ}({\bf A,G}')$
  \State  \Return $\textsc{SelectiveRewire}$
    \EndFunction
  \end{algorithmic}
  \end{algorithm}

\noindent In order to disentangle both effects we now introduce { Algorithm} \ref{Algo4}, which applies both for constructing null models for mediation (M) and suppression (S). To construct a surrogate where all suppression has been removed, starting from $\cal A,B,C$, we perform a selective rewiring in $\cal B$, where only those links in $\cal B$ which are also present in $\cal A$ but not in $\cal C$ (or that also appear in $\cal C$ but not in $\cal A$) are rewired randomly. Similarly, to construct a surrogate where all mediation has been removed, starting from $\cal A,B,C$, we perform a selective rewiring in $\cal B$, where only those links in $\cal B$ which are also present in $\cal A$ and in $\cal C$ are rewired randomly.\\
We then compute again the net Jaccard difference on the rewired versions, which are
are labelled $\Delta_S$ (applied to the case where suppression is removed) and $\Delta_M$ (applied to the case where mediation is removed) respectively. The heuristic is then simple: if there is e.g. hidden suppression in the data (respectively mediation), then $\Delta_S < \Delta$ (respectively $\Delta_M > \Delta$), whereas if such mechanism is absent then $\Delta_S\approx \Delta$ ($\Delta_M\approx\Delta$).\\
Now, we also need to take into account finite-size effects which irremediably add spurious mediation and suppression effects (i.e. triplets of purely random, uncorrelated networks will show small but non-zero mediation and suppression due to chance). To counterbalance such effects, we also proceed to selectively remove suppression and mediation from a completely randomly rewired version of $\cal B$, which we call $\bf B2$, leading to two new indices: $\Delta_{RS}$ and $\Delta_{RM}$. We can finally combine these to produce normalised indices of mediation ($\bar m$) and suppression ($\bar s$) by normalising them dividing  over the maximum possible value of suppression (mediation) attainable by a generative model such as algorithm $3$ ($2$) to the triplet of networks, i.e.
$$ {\bar m}= \frac{\Delta_{S}-\Delta_{\text{R,S}}}{M_{max}}, \ \ {\bar s}= \frac{\Delta_{M}-\Delta_{\text{R,M}}}{S_{max}}.$$
Accordingly, the role that $\cal C$ plays in the relation of $\cal A$  and $\cal B$ is described by the duple $(\bar m, \bar s)$.\\
Additionally, a significance value for these indices could be defined as:
$$\sigma_{\text{\{S,M\}}} = \left\lvert \frac{\Delta_{\{S,M\}}-\Delta_{\text{R,\{S,M\}}}}{\text{std}(\Delta_{\text{R,\{S,M\}}})} \right\rvert.$$
It is worth to stress that for large networks the finite size effects become less common, and $\Delta_{\text{R,\{S,M\}}}$ will tend to zero. In the same way any value of suppression and mediation will be highly significant.\\

\begin{figure}[!htb]
   \centering
   \includegraphics[width=0.85\textwidth]{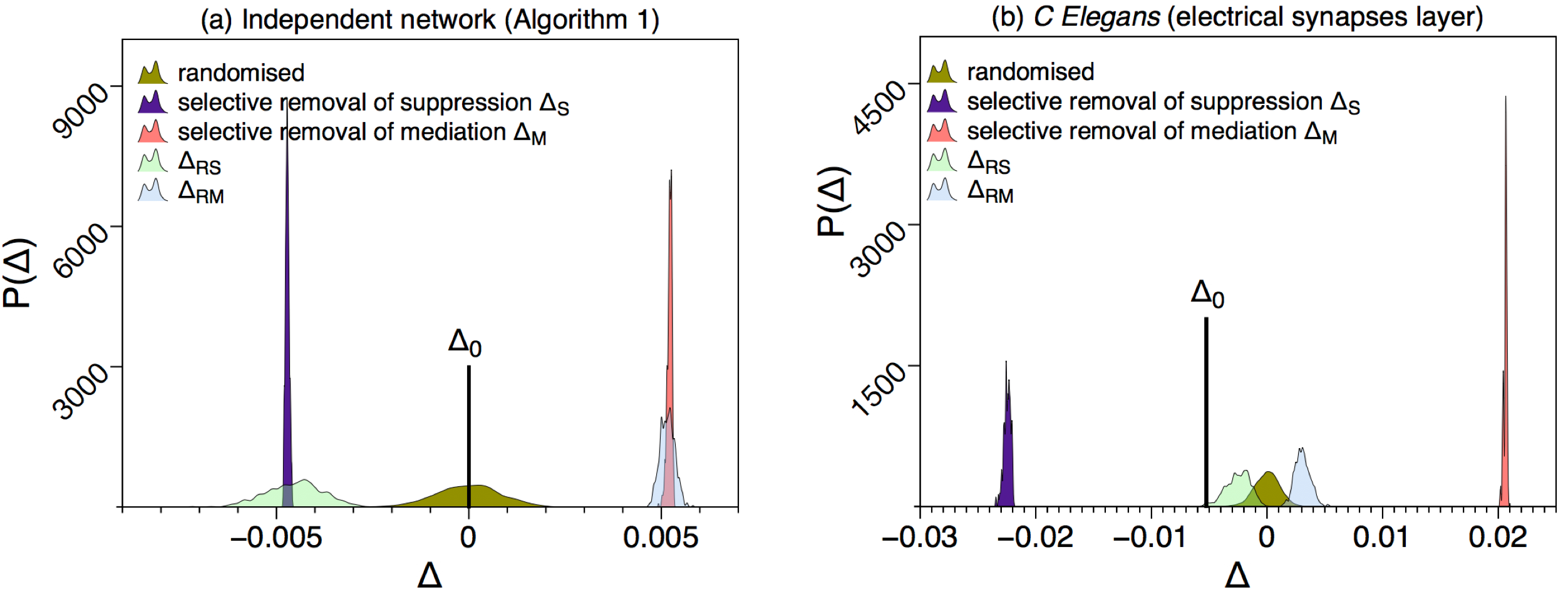}
   \caption{{\bf Illustration of Algorithm 4. } We illustrate the procedure of how to disentangle mediation and suppression in two examples. (a) Selective rewiring applied on a triplet of independent networks generated by Algorithm 1, with $N=300$ nodes and $p=0.3$. The original value of the net Jaccard difference (close to zero) is denoted $\Delta_{0}$. Each selective rewiring is repeated 500 times, and histograms associated to each process are built. The selective removal of suppression and mediation yields only a marginal change in $\Delta$, similar to the   one performed on a randomised version, what indicates that the amount of suppression and mediation in this configuration is residual and only due to finite size effects, as expected since $\cal C$ is independent of $\cal A, B$ by construction.
   (b) Similar to panel (a) but applied on the C Elegans triplet, where network $\cal C$ is assigned to the electrical layer. The actual value $\Delta_{0}<0$, initially suggesting mediation. The selective removal of suppression (mediation) significantly push the histograms towards more negative (positive) values of $\Delta$ --much more than such selective removal performed on a randomisation-- suggesting that there exist a significant amount of mediation and suppression. All histograms are built 
   from $500$ independent rewiring realisations. 
   }
   \label{fig:random_hists}
\end{figure}

\noindent For illustration, in the left panel of Fig.\ref{fig:random_hists} we show the effects of the sequence of selective rewirings on $\Delta$ applied to a particular example of three independent, Erdos-Renyi networks with $N=300$ nodes and wiring probability $p=0.3$ (i.e., Algorithm 1). The original value of $\Delta$ is very close to zero, as well as the ones obtained from a full randomisation of $\cal B$.  Since in this example the networks are independent, any mediation or suppression is only a spurious residual due to finite size effects, thus this residual is flagged out in similar terms by a selective rewiring on the actual network $\cal B$ ($\Delta_X$) or on its full randomisation ($\Delta_{RX}$), hence the violet and green histograms overlap, and similarly the pink and pale blue ones also overlap.
As an additional illustration, we applied the sequence of selective rewirings on the results of the interpolating model. Results are shown in the right panel of Fig.\ref{fig:interpol}.

\section{Empirical networks}

We now turn to real-world networks and consider four types of 3-layer multiplex networks, including (i) different modes of social interaction in Twitter during the 2014's New York City Climate March (NYC), (ii) different types of social interaction --proximity, phone call/text  message, Facebook-- as collected in Denmark (Copenhagen), (iii) different interpersonal relations inside a corporation (Lazega law firm) and (iv) different synaptic junctions in a brain network (C Elegans), see appendix for details.\\
To begin with, in Fig.\ref{fig:J} we confirm that, with the exception of the pair NYC Retweets vs Replies, all other possible pairs of layers in the four examples are indeed genuinely related --i.e., showing substantially more similarity than a null model--. In  each case we plot  $\text{NJ}({\cal A,B})$ (blue bars) and as a reference, as black lines we also plot the average result of $\text{NJ}({\cal A,B})$ ($\pm$ one standard deviation) after $\cal A,B$ have been appropriately randomised. We confirm that the similarity between each pair of networks is not the result of a finite-size effect and thus exploring the role of a third network ($\cal C$) is justified.\\
We then turn to analyse the role of $\cal C$. For illustration, the whole selective rewiring process described in Algorithm 4 is depicted with detail for a specific example (the case of the C Elegans multiplex where we explore the role that the electrical synapses layer play in the  relation between the monadic and the polyadic layers) in the right panel of figure \ref{fig:random_hists}. We provide the original value $\Delta_{0}$, and the distributions of the $\Delta$ values obtained after each of the rewiring procedure, concluding  that this network indeed shows non-negligible mediation and suppression effects.\\

\begin{figure}[!htb]
   \centering
   \includegraphics[width=0.75\textwidth]{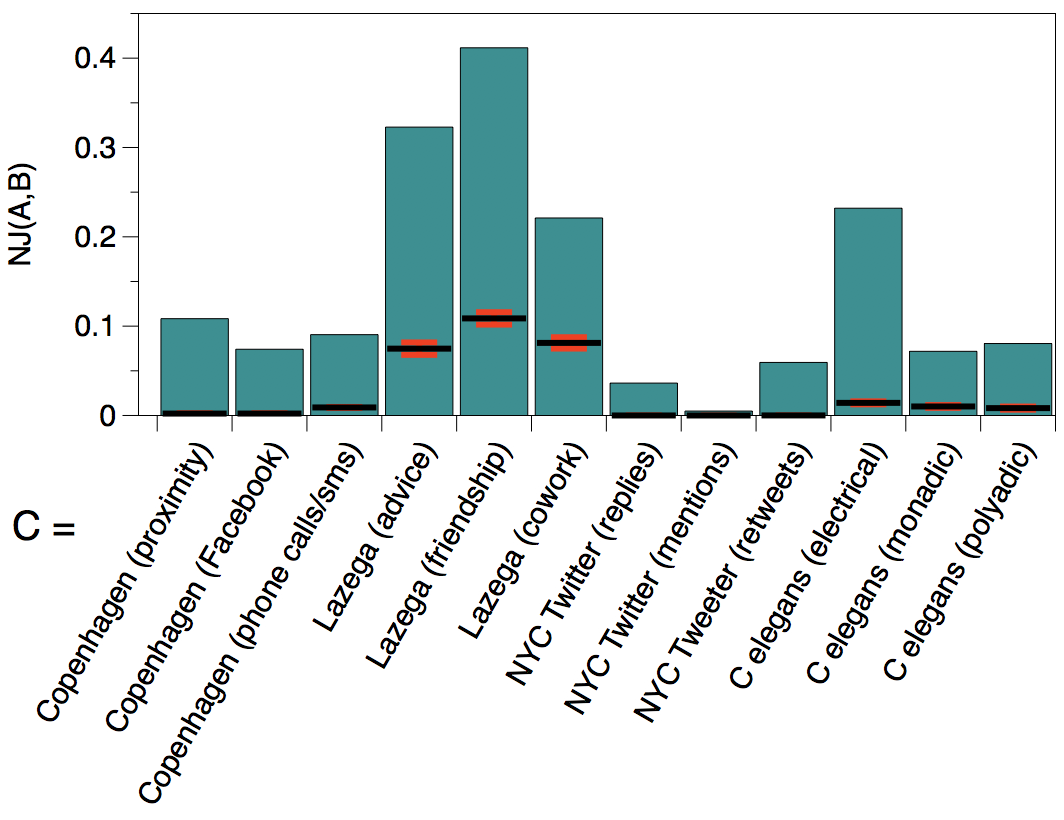} 
   \caption{{\bf Similarity between pairs of real-world networks. } Values of NJ$({\cal A,B})$ computed on the four empirical multiplex networks considered (for each multiplex, we consider the three pair permutations). As a reference, black horizontal lines display NJ$_{\text{null}}({\cal A,B})$ which computes the average over several randomisations of layers $\cal A, B$ (red lines correspond to $\pm$ one  standard deviation. We conclude that all pair of layers are genuinely related with the possible exception of the  pair NYC replies vs retweet.}
   \label{fig:J}
\end{figure}

\begin{figure}[!htb]
   \centering
   \includegraphics[width=0.85\textwidth]{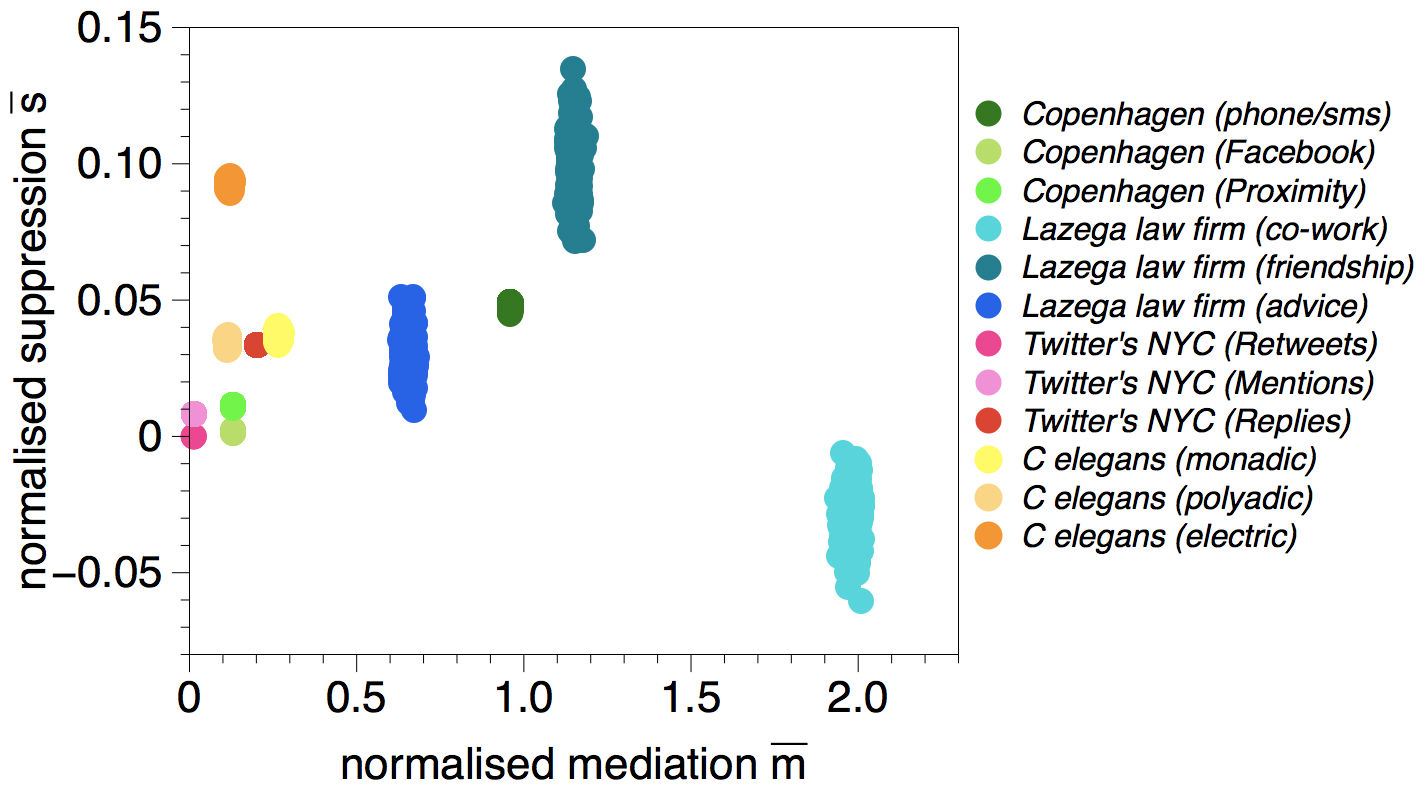} 
   \caption{Normalised indices of mediation and suppression for four real-world multiplex networks. For each multiplex we permute the role of $\cal C$ across the three layers. Dot clouds are the result of repeating the rewiring procedures 500 realisations (for multiplexes with a large number of nodes, the figure displays very little dispersion, and the cloud is only perceptible for the Lazega triplet which is indeed the smallest multiplex).}
   \label{fig:med_syn_2d}
\end{figure}

\noindent The normalized indices of mediation and suppression for the rest of permutations in all the real-world multiplex networks are reported in figure \ref{fig:med_syn_2d}. The first thing we can observe is that overall there is substantially more mediation than suppression, although we also observe the latter mechanism. All effects are statistically significant ($\sigma\gg1$ in every case, data not shown) except for the suppression in the Lazega advice layer and the NYC Retweet layer, where $\sigma \approx 2$ in both cases.\\
In the case of the Copenhagen multiplex, the only layer which evidences a significant role in the relation of the other two is the phone/sms layer, which we show displays both mediation and suppression effects, although the former is notably stronger. For the proximity network we considered averaged values over the whole four weeks, and used an adjacency matrix of a density comparable to the one of Facebook links, corresponding to the closest proximity range. Also the phone network was built irrespective of the timing of the interaction.\\
In the case of the Lazega law firm, all three layers display very high mediation, but such effect is notably stronger for the co-working network, i.e. within this firm dyadic friendships are related to the dyadic advisory relations, and this is mediated by the fact that these are co-working. Only the friendship layer displays a suppressor effect (the one played by the advice layer is non-significant), i.e. pairs of individuals that are not co-working can have an advisory relationship (or otherwise) {\it because} they are friends, or pairs of co-working will also have an advisory relationship  without the needs of them being friends.\\
In the case of the Twitter triplet (NYC), only the Replies network shows a mediating effect.\\
Finally, in the nervous system multiplex (C Elegans) we can see that all layers display some amount of mediation and suppression. The electrical synapses layer is the one displaying a stronger suppressing effect, whereas the monadic chemical layer is the one that displays a larger mediation role.  The increased suppression role of the layer of electrical synapses reflects the evidence that chemical and electrical synapses closely interact and serve related functions \cite{pereda_celegans}, so that when the either of the chemical layers is taken as $\cal{C}$ the presence of the other chemical layer accounts for a reduced suppression/mediation.\\


 \begin{figure}[!htb]
   \centering
   \includegraphics[width=0.75\textwidth]{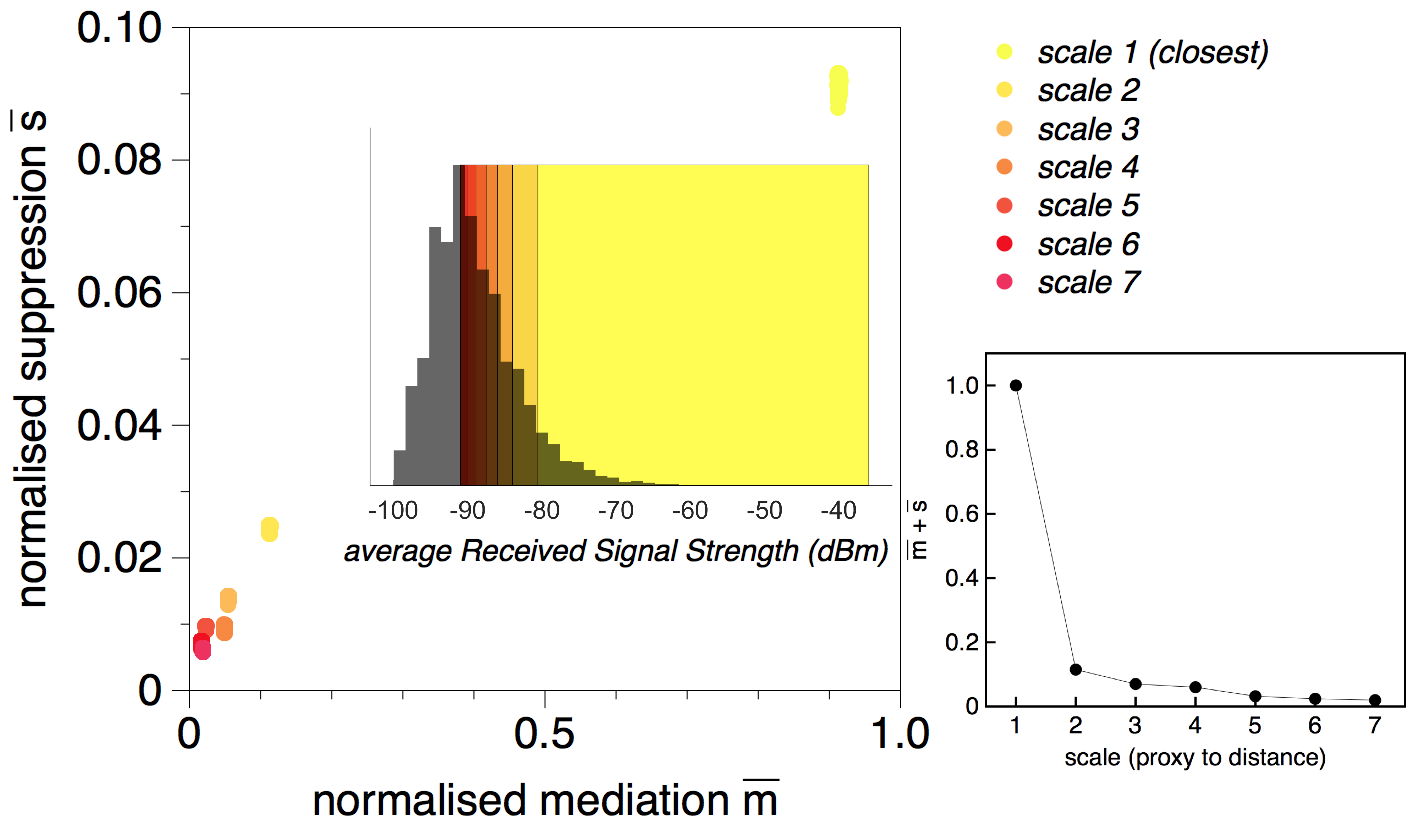} 
   \caption{{\bf Modulating mediation. } (Outer panel) $(\bar m,\bar s)$-plane for the role of the proximity network with respect to the relation of the Facebook and the phone/sms network (Copenhagen multiplex), for different proximity networks reconstructed by  changing the spatial scale over which links are defined. The network displays a very clear mediating effect only when links represent close physical proximity (yellow), for other spatial scales such effect is lost. (Inset  panel) Link average weight distribution (Received Signal Strength (RSSI)) is depicted in black. According to this distribution, we build seven non-overlapping RSSI windows. RSSI is inversely related to distance (the larger the signal strength, the closer two nodes are) so these windows represent  different spatial scales. The right panel describes $\bar m + \bar s$ for each spatial scale, further emphasising  that only in the closer spatial scale the proximity network is playing a strong mediating role.}
   \label{fig:copmod}
\end{figure}

\noindent As a final analysis, and in order to show how suppression and mediation can be functionally modulated within a particular real-world example, we examine the role played by the Proximity layer in the relation between the Facebook and phone calls/sms layers when such layer is systematically varied. In this multiplex, the proximity network is originally reconstructed using Bluethooth signal strength between participants, by assigning a link between each pair of nodes whose relative Bluetooth strength, averaged over the whole period of the recording (four weeks) belongs to a given range. In order to build different proximity networks (each of them accounting for a different  spatial scale) and at the  same time keeping the edge density constant, we build non-overlapping Bluetooth intensity ranges by taking into account  the original Bluetooth intensity distribution  (see the inset of Fig.\ref{fig:copmod}). In this way, ranges are non-uniform but the number of edges in each of this range is the same, hence the resulting proximity networks have  all the  same edge density while describing different scales of physical proximity. The intuition is that only the smaller scale is a meaningful proximity network, and for larger spatial scales the resulting proximity networks do not really imply any real interaction between the nodes. Then, for each resulting proximity network, we estimate the role it plays in the relation between  the other two networks and plot it in the $(\bar m, \bar s)$-plane. Results are shown in the outer panel of Fig.\ref{fig:copmod}.
For proximity networks describing large spatial scales, the network is essentially independent of the other two, and it is only when the proximity network captures smaller spatial scale (i.e. when links describe real physical proximity) that an indirect effect (notably, mediation) gets amplified. 

\section{Discussion}
  
In this paper we have proposed a simple strategy to assess the role that a given network might play 
in shaping the relation between other two networks, thus enlarging the paradigm of network similarity beyond the classical pairwise comparison. This approach is aligned to a recent endeavour that aims at going beyond dyadic interactions
in the characterisation of complex systems \cite{physrep}, and takes inspiration from the causal mediation literature \cite{mackinnon, lizierIID}. We make use of a set-theoretic approach to define a similarity metric between a pair of networks and to further explore if such relation is independent of, mediated or suppressed by a third network which might be hidden. We introduce simple generative models that, we prove, produce pure mediation and suppression. We then explore the coexistence between mediation and suppression and develop a procedure to disentangle both indirect effects. The whole methodology is subsequently applied to a range of real-world, 3-layer multiplex networks, and we unveil previously unnoticed mediation and suppression effects in social and brain networks.\\
We hope this work sparks further research in several areas. First, the simplicity and tractability of the approach makes it easily applicable across the disciplines. Second, our approach can be readily generalised to consider not just isolated triplets of networks. Indeed, one can sequentially apply this protocol to a multiplex network of arbitrary number of layers, or to a temporal network, and accordingly derive concepts of causality and directionality in this context.

\section*{Acknowledgements} LL acknowledges funding from EPSRC EP/P01660X/1. SS was supported by MIUR project PRIN 2017WZFTZP “Stochastic forecasting in complex systems”. DM was supported by the Belgian Embassy in the United Kingdom through the Belgian Chair at the University of London and by the Flemish Fund for Research (FWO) through a sabbatical bench fee.

\section*{Appendix: Proofs of theorems}
\subsection*{Proof of theorem 1}
\proof{Since $\bf A$, $\bf B$ and $\bf C$ are ER(p), all links are independent. We then use expected values, which should be representative as $N$ is large, to directly compute the expected size of the different sets in eq. 
\ref{eq:Jaccard}. With a total of $\ell :=L(L-1)/2$ independent trials, it is easy to see that the expected sizes are
$$\mathbb{E}(|a|)=\mathbb{E}(|b|)=2\ell p,\ \mathbb{E}(|a\cap b|)=2\ell p^2.$$
Since $|a\cup b|=|a| + |b| - |a\cap b|$ we also have
$$\mathbb{E}(|a\cup b|)=2\ell (2p- p^2),$$
such that
$$\mathbb{E}(J(A,B))=\frac{p}{2-p}$$
Consider now $J_p(A,B|C)$. On the one hand, we have $(a\cap b)\setminus c=(a\cap b) \cap c'$, where $c'$ is the complement of $c$. Since all networks are independent, 
$$\mathbb{E}(|(a\cap b)\setminus c|) = 2\ell p^2(1-p)$$
On the other hand, we have
$$\mathbb{E}(|(a\cup b)\setminus c|)=2\ell \cdot (1-p) \cdot \underbrace{Prob(a\cup b)}_{p(2-p)}=2\ell\cdot(1-p)\cdot(2p-p^2),$$
so that altogether $\mathbb{E}(J_p(A,B|C))=\mathbb{E}(J(A,B))=\frac{p}{2-p}$, and $\mathbb{E}(\Delta)=\mathbb{E}(J_p(A,B|C)) - \mathbb{E}(J(A,B))=0$. \qed
}

\subsection*{Proof of theorem 2}
\proof{In this case the proof uses basic arguments of set theory. We aim to prove
that $J_p(A,B|C)<J(A,B)$, i.e.
$$\frac{|(a\cap b)\setminus c|}{|(a\cup b)\setminus c|} < \frac{|a\cap b|}{|a \cup b|}$$
Let us define the residual sets $R_a=a\setminus c$, $R_b := b\setminus c$, and the residuals' intersection $R_i = R_a \cap R_b$ and union $R_u=R_a \cup R_b$. $R_a$ and $R_b$ are clearly non-empty according to Algorithm 2. $R_i$ is not guaranteed to be non-empty as $A$ and $B$ are probabilistic, so we need to assume in what follows that $R_i \neq \O$. $R_u$ is trivially non-empty according to Algorithm 2.\\
Since for any three sets $x$, $y$ and $z$ we have that $(x\cap y)\setminus z= (x\setminus z)\cap(y\setminus z)$ and $(x\cup y)\setminus z= (x\setminus z)\cup(y\setminus z)$, we also have that $|(a\cap b)\setminus c|=|R_i|$ and $|(a\cup b)\setminus c|=|R_u|$. Incidentally, our previous assumption can now be easily interpreted: we need to assume that networks A and B share at least one edge besides the common edges shared with C, that's why this assumption is indeed stated in the theorem.\\
Now, since in this case $c \subset (a\cap b)$ it is easy to see that $a \cap b = c \cup R_i$ and $a\cup b = c\cup R_u$.
Also by construction we have $c \cap R_i = \O$ and therefore $|a \cap b| = |c| + |R_i|$. Similarly, also by construction $c \cap R_u = \O$ and therefore $|a \cup b| = |c \cup R_u|= |c| + |R_u|$. Therefore we aim to prove that 

$$\frac{|R_i|}{|R_u|}<\frac{|c| + |R_i|}{|c| + |R_u|}$$

Since $R_i$ and $R_u$ are respectively the intersection and the union of two sets, it trivially follows that $|R_i|\leq |R_u|$, where inequality only saturates when $a=b$, and otherwise is strict. Let us indeed assume $A\neq B$, thus enforcing the strict inequality. Rearranging terms:
\begin{eqnarray}
&& |R_i|< |R_u| \nonumber \\
\iff && \frac{1}{|R_i|} > \frac{1}{|R_u|}\nonumber \\
\iff && \frac{|c|}{|R_i|} > \frac{|c|}{|R_u|}, \ (|c|>0)\nonumber \\
\iff && 1+\frac{|c|}{|R_i|} > 1+\frac{|c|}{|R_u|} \nonumber \\
\iff && \frac{|R_i|+|c|}{|R_i|} > \frac{|R_u|+|c|}{|R_u|} \nonumber \\
\iff && \frac{|R_i|}{|R_u|}< \frac{|c| + |R_i|}{|c| + |R_u|}
\end{eqnarray}
\qed}

\subsection*{Proof of theorem 3}
\proof{
The first thing to observe is that in {\it Algorithm 3} the condition $c \subset (a\cap b)$ is not met, therefore theorem \ref{thm:mediated} does not hold in this case. 
Let us now define the following sets:\\
Let $a_p\subset a$ be the subset of edges in $a$ such that $a_p \cap c=\O$. The elements of this subset will be in $b$ with probability $q$, hence on average $|a_p|q$ edges of $a_p$ will be in $b$.\\
Let $c_p\subset b$ be the subset of edges in $c$ such that $c_p \cap a=\O$. The elements of this subset will be in $b$ with probability $q$, hence on average $|c_p|q$ edges of $c_p$ will be in $b$.\\
Let $r\subset b$ be the subset of edges in $b$ which are neither in $a$ nor in $c$, i.e. $r\cap (a\cup c)=\O$.\\
By symmetry, we have $\mathbb{E}(|a_p|)=\mathbb{E}(|c_p|)$. According to Algorithm 3, we have\\
$$\mathbb{E}(|(a\cap b)\setminus c|) =  \mathbb{E}(|(a\cap b))=\mathbb{E}(|a_p|) q,$$
$$\mathbb{E}(|(a\cup b)\setminus c|)= \mathbb{E}(|a_p|) + \mathbb{E}(|r|),$$
$$\mathbb{E}(|(a\cup b))=\mathbb{E}(|a_p|) + \mathbb{E}(|c_p|)q + \mathbb{E}(|r|), $$
and thus
$$\mathbb{E}(J_p(A,B|C))=\frac{\mathbb{E}(|a_p|) q}{\mathbb{E}(|a_p|) + \mathbb{E}(|r|)},$$
whereas
$$\mathbb{E}(J(A,B))= \frac{\mathbb{E}(|a_p|) q}{\mathbb{E}(|a_p|) + \mathbb{E}(|c_p|)q + \mathbb{E}(|r|)}.$$
Therefore as long as $\mathbb{E}(|c_p|)q>0$, we have $\mathbb{E}(J_p(A,B|C)) > \mathbb{E}(J(A,B))$, yielding $\mathbb{E}(\Delta)>0$. Since A and C are independent, $c_p$ is not empty with large probability, so given a value of $q$,  for sufficiently large $N$ this condition is always met. \qed
}

\section*{Appendix: empirical networks}

\begin{table}[htb]
    \centering
\begin{tabular}{ |p{2cm}||p{1.99cm}|p{1.35cm}|p{2.5cm}|p{2.3cm}|p{2.cm}|  }
 \hline
 \multicolumn{6}{|c|}{Summary of empirical networks} \\
 \hline
 {\bf Triplet} & {\bf Type} & $N$ &{\bf Network} $\#1$&{\bf Network} $\#2$&{\bf Network} $\#3$\\
 \hline
C. Elegans& brain & 279 & monadic (1639 edges) & polyadic (3193 edges) & electrical (1031 edges) \\
 \hline
NYC& Twitter &102439 & retweet (213754 edges) & mentions (131679 edges) & replies (8062 edges)\\  
\hline
 Lazega law firm & social (offline) & 71 & cowork (892 edges) & friendship (575 edges) & advice (1104 edges)\\
 \hline
Copenhagen & social (offline/online) & 751 & proximity (13020 edges) & facebook (12847 edges) & calls/sms (1760 edges)\\
 \hline
\end{tabular}
\caption{{\bf Summary of network specificities.} The first three examples are multiplex networks collected from \url{comunelab.fbk.eu/data.php}, whereas the fourth one is collected from \url{icon.colorado.edu}. The C Elegans multiplex describes the Caenorhabditis elegans connectome, where layers correspond to different synaptic junctions:  chemical monadic ("MonoSyn"), polyadic ("PolySyn") and electrical ("ElectrJ") \cite{celegans, muxviz}.
The NYC multiplex describes Twitter activity during an exceptional event, the NYC Climate March in 2014, and layers correspond to retweet, mentions and replies \cite{nyc}. The Lazega law firm depicts three kinds of social relationships (Co-work, Friendship and advice) between partners and associates of a corporate law partnership \cite{lazega1,lazega2}.
Finally  the Copenhagen multiplex describes social interaction in three layers corresponding to phone calls and text messages (merged), Facebook friendships, and proximity as measured with strength of Bluetooth signal \cite{copenhagen}. 
}
\label{tab:nets}
\end{table}

\bibliography{apssamp}

\begin{thebibliography}{10}

\bibitem{latora} V. Latora, V. Nicosia, G. Russo, Complex Networks: principles and applications (Cambridge University Press, 2017).

\bibitem{distance0} S.V.N. Vishwanathan, N.N. Schraudolph, R. Kondor, R. and K.M. Borgwardt. Graph kernels. The Journal of Machine Learning Research, 11 (2010)  pp.1201-1242.  
\bibitem{distance1} T.A. Schieber, L. Carpi, A. Diaz-Guilera, P.M. Pardalos, C. Masoller, and M.G. Ravetti,  Quantification of network structural dissimilarities. {\it Nature Communications} 8 (1) (2017).
\bibitem{distance2} M. De Domenico,  and J. Biamonte, Spectral entropies as information-theoretic tools for complex network comparison. Physical Review X, 6(4), p.041062 (2016).
\bibitem{distance3} A. Santoro, V. Nicosia, "Algorithmic complexity of multiplex networks", Phys. Rev. X 10, 021069 (2020).
\bibitem{distance4} M. De Domenico, V. Nicosia, A. Arenas, V. Latora, Structural reducibility of multilayer networks, {\it Nature Communications} 6, 6864 (2015)
\bibitem{distance5}  V. Nicosia, M. De Domenico, V. Latora. Characteristic exponents of complex networks, {\it EPL} 106, 58005 (2014).
\bibitem{distance6} L. Lacasa, J. Gomez-Garde\~nes, Correlation dimension of complex networks, {\it Physical Review Letters} (2013).

\bibitem{toobig} M. Lin, H.C. Lucas Jr, and G. Shmueli. Research Commentary—Too Big to Fail: Large Samples and the p-Value Problem. {\it Information Systems Research} 24:4, 906-917 (2017).

 
\bibitem{physrep} F. Battiston, G. Cencetti, I. Iacopini, V. Latora, M. Lucas, A. Patania, J-G Young, and G. Petri. Networks beyond pairwise interactions: structure and dynamics. {\it Physics Reports} (2020).
\bibitem{mackinnon} D.P. MacKinnon, J.L. Krull, and C.M. Lockwood. Equivalence of the Mediation, Confounding and Suppression Effect, {\it Prev Sci.} (2000).
\bibitem{lizierIID} J.T. Lizier, N. Bertschinger,J. Jost, and M. Wibral. Information Decomposition of Target Effects from Multi-Source Interactions: Perspectives on Previous, Current and Future Work {\it Entropy} 20(4), 307 (2018).

\bibitem{celegans}  B.L. Chen, D.H. Hall, and D.B. Chklovskii. Wiring optimization can relate neuronal structure and function {\it PNAS} 103 (12) 4723–4728 (2006)

\bibitem{pereda_celegans} A.E. Pereda. Electrical synapses and their functional interactions with chemical synapses. {\it Nat Rev Neurosci.} 5(4):250-63 (2014)

\bibitem{muxviz} M. De Domenico, M.A. Porter, and A. Arenas. MuxViz: A Tool for Multilayer Analysis and Visualization of Networks. {\it Journal of Complex Networks} 3 (2) 159-176 (2015)

\bibitem{lazega1} E. Lazega. The Collegial Phenomenon: The Social Mechanisms of Cooperation Among Peers in a Corporate Law Partnership. Oxford University Press (2001)

\bibitem{lazega2} T.A.B. Snijders, P.E. Pattison, G.L. Robins, and M.S. Handcock. New specifications for exponential random graph models.{\it Sociological Methodology} 99-153 (2006).

\bibitem{nyc} E. Omodei, M. De Domenico, A. Arenas. Characterizing interactions in online social networks during exceptional events. {\it Front. Phys}. 3, 59 (2015)

\bibitem{copenhagen} P. Sapiezynski, A. Stopczynski, D.D. Lassen, and S. Lehmann.  Interaction data from the Copenhagen Networks Study. {\it Scientific Data} 6, 315 (2019) 


\end{thebibliography}

\end{document}